\documentclass[aps,preprint,prb]{revtex4}
\usepackage{graphicx}
\newcommand{\E}{\mathbf{E}}
\newcommand{\n}{\mathbf{n}}
\newcommand{\inp}{\text{in}}
\newcommand{\out}{\text{out}}
\begin{document}
\title{Pancharatnam-Berry phase optical elements for wavefront shaping in the visible domain:
switchable helical modes generation}
\author{L. Marrucci}
\email{lorenzo.marrucci@na.infn.it}
\author{C. Manzo}
\author{D. Paparo}
\affiliation{CNR-INFM Coherentia and Dipartimento di Scienze
Fisiche, Universit\`{a} di Napoli ``Federico II'', Compl.\ di Monte
S.Angelo, v.\ Cintia, 80126 Napoli, Italy}
\date{January 19, 2006}
\begin{abstract}
We report the realization of a Pancharatnam-Berry phase optical
element [Z. Bomzon, G. Biener, V. Kleiner, and E. Hasman, Opt. Lett.
\textbf{27}, 1141 (2002)] for wavefront shaping working in the
visible spectral domain, based on patterned liquid crystal
technology. This device generates helical modes of visible light
with the possibility of electro-optically switching between opposite
helicities by controlling the handedness of the input circular
polarization. By cascading this approach, fast switching among
multiple wavefront helicities can be achieved, with potential
applications to multi-state optical information encoding. The
approach demonstrated here can be generalized to other
polarization-controlled devices for wavefront shaping, such as
switchable lenses, beam-splitters, and holographic elements.
\end{abstract}
\maketitle

When the polarization of an electromagnetic wave undergoes a
continuous sequence of transformations following a closed path in
the space of polarization states (e.g., the Poincar\'e sphere), the
wave acquires a phase shift, known as Pancharatnam-Berry phase, that
is determined only by the geometry of the polarization
path.\cite{pancharatnam56,berry87} By the same principle, if a wave
is subjected to transversely inhomogeneous polarization
transformations with a homogeneous initial and final polarization
state, the associated inhomogeneous geometrical phases will induce
an overall wavefront reshaping. This approach to wavefront shaping
is fundamentally different from the usual optical-path-length
approaches of standard lenses, curved mirrors, and gradient-index
(GRIN) elements. It is also conceptually different from holographic
approaches, although the two are related, as we will discuss further
below. The realization of so-called \textit{Pancharatnam-Berry phase
optical elements} (PBOE) for wavefront shaping has been proposed
only recently,\cite{bhandari97,bomzon02} and it has been
experimentally demonstrated only in the mid-infrared domain, using
subwavelengths inhomogeneous gratings to manipulate the
polarization.\cite{bomzon02,biener02,hasman02,hasman03} An
additional general feature of PBOE's is that they are
\textit{polarization-controlled}, i.e. different input polarizations
will give rise to different wavefront shaping in the same PBOE
element. Since the polarization can be electro-optically switched at
a high rate, PBOE's allow a very fast control of the generated
wavefront. This polarization multiplexing is limited to a finite set
of predefined wavefronts, so PBOE's cannot compete with spatial
light modulators in terms of flexibility, but they will be much
faster and cheaper.

To be more specific, let us consider a PBOE made as a single
(uniaxial) birefringent plate having a homogeneous phase retardation
of $\pi$ (half-wave PBOE) for light propagation in the longitudinal
$z$ direction but a transversely inhomogeneous optical axis
$\n(x,y)$, lying in the $xy$ plane. To analyze the effect of this
element on the optical field, it is convenient to adopt the Jones
formalism. Let $\alpha(x,y)$ be the angle between $\n(x,y)$ and a
fixed axis $x$. The Jones matrix $\mathbf{M}$ to be applied on the
field at each transverse position $x,y$ is the following:
\begin{equation}\label{qjones}
\mathbf{M}=\mathbf{R}(-\alpha)\! \left(\!
  \begin{array}{cc}
    1 & 0 \\
    0 & -1 \\
  \end{array}\!
\right)\! \mathbf{R}(\alpha) =\!
                \left(\!
                   \begin{array}{cc}
                     \cos2\alpha & \sin2\alpha \\
                     \sin2\alpha & -\cos2\alpha \\
                   \end{array}\!
                 \right),
\end{equation}
where $\mathbf{R}(\alpha)$ is the two-dimensional rotation matrix by
angle $\alpha$. An input left-circular polarized plane wave,
described by the Jones electric-field vector
$\E_{\inp}=E_0\times[1,i]$, will be transformed by this element into
the following field (up to an overall phase):
\begin{equation}\label{outfield}
\E_{\out}=\mathbf{M}\cdot\E_{\inp} = E_0e^{i2\alpha(x,y)}\left[
                                                            \begin{array}{c}
                                                              1 \\
                                                              -i \\
                                                            \end{array}
                                                          \right].
\end{equation}
It is seen that the output wave is uniformly right-circular
polarized, but its wavefront has acquired a nonuniform phase
retardation $\Delta\Phi(x,y)=2\alpha(x,y)$. If the input light is
right-circular polarized, it is easy to verify that the wavefront is
the conjugate one, i.e. $\Delta\Phi(x,y)=-2\alpha(x,y)$.

\begin{figure}[h]
\includegraphics[angle=0, width=0.8\textwidth]{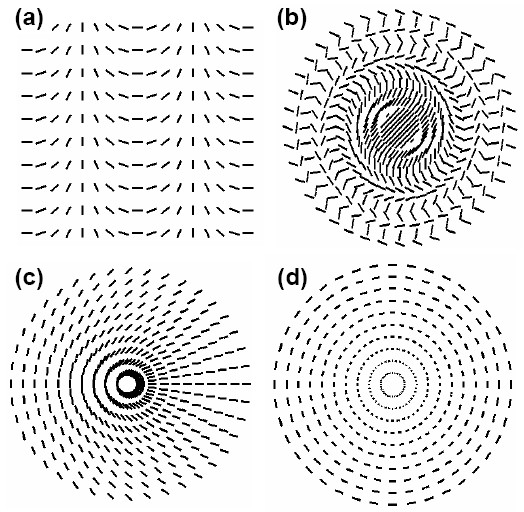}
\caption{Examples of half-wave PBOE geometries. Dashes indicate
local optical axis direction. (a) PBOE behaving as a
circular-polarizing beam-splitter or switch; (b) PBOE behaving as a
polarization-dependent lens; (c) $q$-plate PBOE with $q=1/2$ and
$\alpha_0=0$, generating helical modes of order ${\pm}1$; (d)
$q$-plate PBOE with $q=1$ and $\alpha_0=\pi/2$, generating helical
modes of order ${\pm}2$.} \label{pboegeofig}
\end{figure}

To appreciate the possible applications of these devices, consider,
for example, a half-wave PBOE having a polarization-grating geometry
as that shown in Fig.~\ref{pboegeofig}a. This device will function
as a circular-polarizing beam-splitter or as polarization-controlled
optical switch.\cite{hasman02} A PBOE lens can instead be obtained
with an optical axis geometry given by $\alpha{\propto}r^2$, where
$r$ is the radial coordinate in the $xy$ plane, as that shown in
Fig.~\ref{pboegeofig}b. This element will be focusing or defocusing,
depending on the input circular polarization
handedness.\cite{bhandari97,hasman03}

Let us consider now a PBOE geometry given by
$\alpha=q\varphi+\alpha_0$, where $\varphi$ is the azimuthal angle
in the $xy$ plane, and $q$ and $\alpha_0$ are two constants. We
further assume that $q$ is an integer or a semi-integer, so that the
optical axis does not have discontinuity lines in the plate, but
only a defect in the center. We will call these devices
``$q$-plates''. Figures \ref{pboegeofig}c and \ref{pboegeofig}d show
examples of these devices for $q=1/2$ and $q=1$, respectively. These
$q$-plates give rise to a wavefront modulation given by
$\Delta\Phi=\pm2q\varphi$, with a sign depending on the input
circular polarization handedness, i.e. they generate helical
wavefronts of order $\pm2q$.\cite{allen92,sundbeck05} Thus far,
$q$-plates have been demonstrated only for the mid-infrared
wavelength of $10.6$ $\mu$m, based on the subwavelength gratings
technology.\cite{biener02,niv05}

\begin{figure}[h]
\includegraphics[angle=0, width=0.8\textwidth]{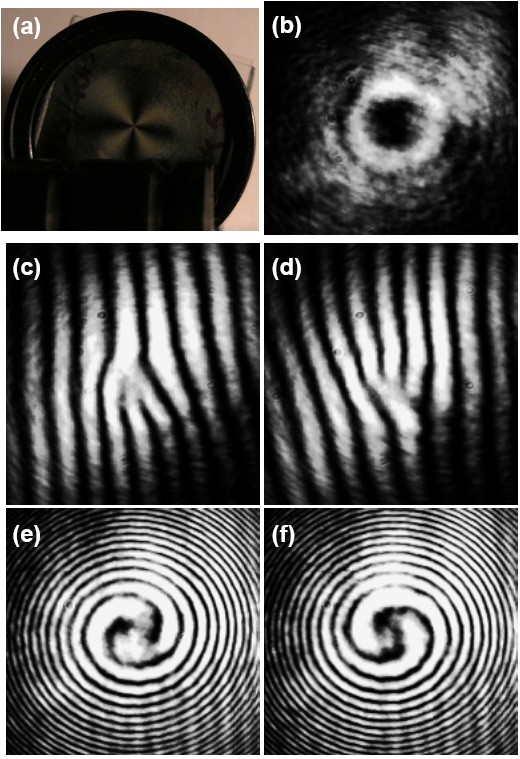}
\caption{Experimental images. (a) A LC $q$-plate held between
crossed polarizers, showing the expected pattern for $q=1$ geometry.
(b) ``Doughnut'' intensity profile of the beam emerging from the
$q$-plate. (c-f) Interference patterns of helical modes generated by
our $q$-plate. (c-d) panels refer to the plane-wave reference
geometry, (e-f) panels to the spherical-wave reference one. Panels
on the left, (c) and (e), are for a left-circular input polarization
and those on the right, (d) and (f), for a right-circular one.}
\label{expfig}
\end{figure}

We manufactured $q=1$ plates working at the visible wavelength
$\lambda=633$ nm based on the patterned liquid crystal (LC)
technology (see, e.g., Refs.\ \onlinecite{varghese04,syed05} and
references therein). Nematic LC planar cells were prepared with a
thickness (about 1 $\mu$m) and a material (E63 from Merck,
Darmstadt, Germany) chosen so as to obtain a birefringence
retardation of approximately a half wave. Before cell assembly, one
of the inner surfaces of the two containing glasses of the cell was
pressed against a piece of fabric kept in continuous rotation. This
``circular rubbing'' procedure leads to a surface easy axis (i.e.
the preferred orientation of LC molecules) having the desired $q=1$
circular-symmetric geometry, as that shown in
Fig.~\ref{pboegeofig}d. The other glass was left unrubbed, for
degenerate planar alignment. To ensure good LC alignment, the cell
was heated above the clearing point and then cooled slowly, keeping
the rubbed surface slightly colder than the unrubbed one. In this
way, nematic order nucleated on the rubbed surface and then extended
to the whole cell. Some cells were prepared with a polyimide coating
for planar alignment, others with bare glass, with comparable
results (although they required different rubbing pressures and
lengths). A photograph of a LC $q$-plate held between crossed
polarizers is shown in Fig.~\ref{expfig}a.

To test the optical effect of a $q$-plate, a circularly-polarized
He-Ne laser beam having a TEM$_{00}$ transverse mode and a
beam-waist radius of about 1 mm was sent through it, taking care of
aligning the beam axis on the $q$-plate center. The intensity
profile of the output beam, shown in Fig.~\ref{expfig}b, has the
``doughnut'' shape expected for a helical mode. However, a complete
test must be based on measuring the beam wavefront shape, rather
than its intensity profile. To this purpose, we inserted the
$q$-plate in the signal arm of a Mach-Zender interferometer based on
the same He-Ne laser source. The input circular-polarization
handedness was selected by properly rotating a quarter-wave plate.
The beam emerging from the $q$-plate was sent through another
quarter-wave plate and a linear polarizer arranged for transmitting
the polarization handedness opposite to the initial one, so as to
eliminate any residual unchanged circular polarization (this step
would be unnecessary for an exact half-wave retardation of the
$q$-plate). The final interference pattern generated after
superposition with the reference was formed directly on the sensing
area of a CCD camera. We used two reference wavefront geometries:
(i) plane-tilted, for which an order-2 helical wavefront will give
rise to a double disclination defect in a otherwise regular
straight-line fringe pattern; (ii) spherical, for which the helical
wavefront will give rise to a double spiral fringe pattern. Figures
\ref{expfig}c-f show the interference patterns we obtained for one
of our cells in these two geometries, respectively for a
left-circular (panels c and e) and right-circular (panels d and f)
input polarization. These results confirm that the wavefront of the
light emerging from our $q$-plate is indeed helical of order $\pm2$,
as expected, with the $\pm$ sign determined by the input
polarization handedness.

\begin{figure}[h]
\includegraphics[width=0.95\textwidth]{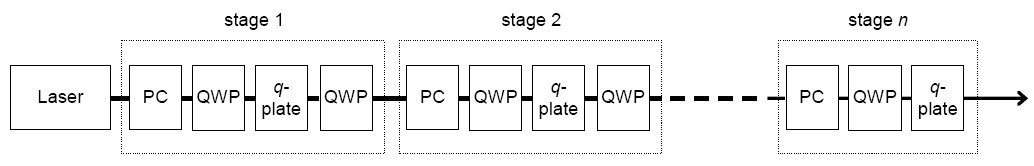}
\caption{A $n$ stages PBOE optical system for generating helical
modes of light having an order $l$ which can be electro-optically
switched in the set $l\in\{-2nq,-2(n-2)q,\ldots,+2(n-2)q,+2nq\}$.
Legend: PC - Pockel cell; QWP - quarter-wave plate.}
\label{cascadefig}
\end{figure}

This polarization-based control of the generated helical wavefront
is a good example of the possible advantages of the PBOE approach to
wavefront shaping. Indeed, all other existing approaches to helical
mode generation (i.e. cylindrical lenses, spiral phase plates, and
holographic methods) have an essentially fixed output. Of course, by
introducing a suitable spatial light modulator, dynamical control
becomes possible, but only at relatively low switching rates. In our
approach, a simple electro-optical control of the input polarization
allows switching of the generated helical mode at very high rate. By
cascading several $q$-plates in series with suitable electro-optic
devices in between, as shown in Fig.~\ref{cascadefig}, one can
obtain fast switching among several different helical orders. This
could be very useful if helical modes are to be used in multi-state
optical information encoding, as recently proposed for classical
communication\cite{gibson04} and for quantum communication and
computation.\cite{leach02,vaziri02}

Although in our proof-of-principle demonstration reported here we
used a method for patterning our LC cell that works only for
circular-symmetric geometries (as in the $q=1$ plate), LC cell
patterning has the potential for obtaining any desired PBOE
geometry. Different approaches, such as
micro-rubbing,\cite{varghese04} masked or holographic
photo-alignment,\cite{schadt96,fan03} and silicon-oxide evaporated
coatings,\cite{chen95} are all suitable.

Finally, we note that the PBOE principle is strictly related to the
so-called \textit{polarization holography} (PH), in which the
holographic material records the information contained in the
optical polarization.\cite{todorov84} Typically, in PH one needs a
light-sensitive polymer that can align its molecular chains parallel
or perpendicular to the polarization direction.\cite{eich87} In
order to memorize a given wavefront in a PH hologram, one must
superimpose the wave carrying that wavefront with a plane-wave
reference, taking care that both waves are circularly polarized,
with opposite handedness. The resulting interference field will have
uniform intensity and it will be everywhere linearly polarized, but
it will have a nonuniform polarization orientation which will be
imprinted in the hologram. When illuminated with a plane wave, this
hologram will reconstruct the recorded wavefront, or its conjugate,
at its $\pm1$ diffraction orders. However, if the hologram is
``developed'' into a inhomogeneous birefringent plate having
half-wave retardation (for example by using the hologram as a
``command'' surface of a LC cell, or if the hologram itself has
sufficient birefringence), the zero diffraction order will vanish
identically and the hologram becomes a PBOE device generating a
single optical output with the recorded wavefront, or its conjugate,
when illuminated with a circularly polarized plane wave.

In conclusion, we have demonstrated a Pancharatnam-Berry phase
optical element working in the visible domain, based on patterned
liquid crystal technology. This device can be used for generating
fast switchable helical modes, with potential applications to
optical information encoding. Some plausible strategies for
generalizing our approach have been discussed.


\newpage
\begin{thebibliography}{20}
\expandafter\ifx\csname
natexlab\endcsname\relax\def\natexlab#1{#1}\fi
\expandafter\ifx\csname bibnamefont\endcsname\relax
  \def\bibnamefont#1{#1}\fi
\expandafter\ifx\csname bibfnamefont\endcsname\relax
  \def\bibfnamefont#1{#1}\fi
\expandafter\ifx\csname citenamefont\endcsname\relax
  \def\citenamefont#1{#1}\fi
\expandafter\ifx\csname url\endcsname\relax
  \def\url#1{\texttt{#1}}\fi
\expandafter\ifx\csname urlprefix\endcsname\relax\def\urlprefix{URL
}\fi \providecommand{\bibinfo}[2]{#2}
\providecommand{\eprint}[2][]{\url{#2}}

\bibitem[{\citenamefont{Pancharatnam}(1956)}]{pancharatnam56}
\bibinfo{author}{\bibfnamefont{S.}~\bibnamefont{Pancharatnam}},
  \bibinfo{journal}{Proc.\ Indian Acad.\ Sci.\ Sect.\ A}
  \textbf{\bibinfo{volume}{44}}, \bibinfo{pages}{247} (\bibinfo{year}{1956}).

\bibitem[{\citenamefont{Berry}(1987)}]{berry87}
\bibinfo{author}{\bibfnamefont{M.~V.} \bibnamefont{Berry}},
  \bibinfo{journal}{J. Mod.\ Opt.} \textbf{\bibinfo{volume}{34}},
  \bibinfo{pages}{1401} (\bibinfo{year}{1987}).

\bibitem[{\citenamefont{Bhandari}(1997)}]{bhandari97}
\bibinfo{author}{\bibfnamefont{R.}~\bibnamefont{Bhandari}},
  \bibinfo{journal}{Phys.\ Rep.} \textbf{\bibinfo{volume}{281}},
  \bibinfo{pages}{1} (\bibinfo{year}{1997}).

\bibitem[{\citenamefont{Bomzon et~al.}(2002)\citenamefont{Bomzon, Biener,
  Kleiner, and Hasman}}]{bomzon02}
\bibinfo{author}{\bibfnamefont{Z.}~\bibnamefont{Bomzon}},
  \bibinfo{author}{\bibfnamefont{G.}~\bibnamefont{Biener}},
  \bibinfo{author}{\bibfnamefont{V.}~\bibnamefont{Kleiner}}, \bibnamefont{and}
  \bibinfo{author}{\bibfnamefont{E.}~\bibnamefont{Hasman}},
  \bibinfo{journal}{Opt.\ Lett.} \textbf{\bibinfo{volume}{27}},
  \bibinfo{pages}{1141} (\bibinfo{year}{2002}).

\bibitem[{\citenamefont{Biener et~al.}(2002)\citenamefont{Biener, Niv, Kleiner,
  and Hasman}}]{biener02}
\bibinfo{author}{\bibfnamefont{G.}~\bibnamefont{Biener}},
  \bibinfo{author}{\bibfnamefont{A.}~\bibnamefont{Niv}},
  \bibinfo{author}{\bibfnamefont{V.}~\bibnamefont{Kleiner}}, \bibnamefont{and}
  \bibinfo{author}{\bibfnamefont{E.}~\bibnamefont{Hasman}},
  \bibinfo{journal}{Opt.\ Lett.} \textbf{\bibinfo{volume}{27}},
  \bibinfo{pages}{1875} (\bibinfo{year}{2002}).

\bibitem[{\citenamefont{Hasman et~al.}(2002)\citenamefont{Hasman, Bomzon, Niv,
  Biener, and Kleiner}}]{hasman02}
\bibinfo{author}{\bibfnamefont{E.}~\bibnamefont{Hasman}},
  \bibinfo{author}{\bibfnamefont{Z.}~\bibnamefont{Bomzon}},
  \bibinfo{author}{\bibfnamefont{A.}~\bibnamefont{Niv}},
  \bibinfo{author}{\bibfnamefont{G.}~\bibnamefont{Biener}}, \bibnamefont{and}
  \bibinfo{author}{\bibfnamefont{V.}~\bibnamefont{Kleiner}},
  \bibinfo{journal}{Opt.\ Commun.} \textbf{\bibinfo{volume}{209}},
  \bibinfo{pages}{45} (\bibinfo{year}{2002}).

\bibitem[{\citenamefont{Hasman et~al.}(2003)\citenamefont{Hasman, Kleiner,
  Biener, and Niv}}]{hasman03}
\bibinfo{author}{\bibfnamefont{E.}~\bibnamefont{Hasman}},
  \bibinfo{author}{\bibfnamefont{V.}~\bibnamefont{Kleiner}},
  \bibinfo{author}{\bibfnamefont{G.}~\bibnamefont{Biener}}, \bibnamefont{and}
  \bibinfo{author}{\bibfnamefont{A.}~\bibnamefont{Niv}},
  \bibinfo{journal}{Appl.\ Phys.\ Lett.} \textbf{\bibinfo{volume}{82}},
  \bibinfo{pages}{328} (\bibinfo{year}{2003}).

\bibitem[{\citenamefont{Allen et~al.}(1992)\citenamefont{Allen, Beijersbergen,
  Spreeuw, and Woerdman}}]{allen92}
\bibinfo{author}{\bibfnamefont{L.}~\bibnamefont{Allen}},
  \bibinfo{author}{\bibfnamefont{M.~W.} \bibnamefont{Beijersbergen}},
  \bibinfo{author}{\bibfnamefont{R.~J.~C.} \bibnamefont{Spreeuw}},
  \bibnamefont{and} \bibinfo{author}{\bibfnamefont{J.~P.}
  \bibnamefont{Woerdman}}, \bibinfo{journal}{Phys.\ Rev.\ A}
  \textbf{\bibinfo{volume}{45}}, \bibinfo{pages}{8185} (\bibinfo{year}{1992}).

\bibitem[{\citenamefont{Sundbeck et~al.}(2005)\citenamefont{Sundbeck, Gruzberg,
  and Grier}}]{sundbeck05}
\bibinfo{author}{\bibfnamefont{S.}~\bibnamefont{Sundbeck}},
  \bibinfo{author}{\bibfnamefont{I.}~\bibnamefont{Gruzberg}}, \bibnamefont{and}
  \bibinfo{author}{\bibfnamefont{D.~G.} \bibnamefont{Grier}},
  \bibinfo{journal}{Opt.\ Lett.} \textbf{\bibinfo{volume}{30}},
  \bibinfo{pages}{477} (\bibinfo{year}{2005}).

\bibitem[{\citenamefont{Niv et~al.}(2005)\citenamefont{Niv, Biener, Kleiner,
  and Hasman}}]{niv05}
\bibinfo{author}{\bibfnamefont{A.}~\bibnamefont{Niv}},
  \bibinfo{author}{\bibfnamefont{G.}~\bibnamefont{Biener}},
  \bibinfo{author}{\bibfnamefont{V.}~\bibnamefont{Kleiner}}, \bibnamefont{and}
  \bibinfo{author}{\bibfnamefont{E.}~\bibnamefont{Hasman}},
  \bibinfo{journal}{Opt.\ Commun.} \textbf{\bibinfo{volume}{251}},
  \bibinfo{pages}{306} (\bibinfo{year}{2005}).

\bibitem[{\citenamefont{Varghese et~al.}(2004)\citenamefont{Varghese, Crawford,
  Bastiaansen, de~Boer, and Broer}}]{varghese04}
\bibinfo{author}{\bibfnamefont{S.}~\bibnamefont{Varghese}},
  \bibinfo{author}{\bibfnamefont{G.~P.} \bibnamefont{Crawford}},
  \bibinfo{author}{\bibfnamefont{C.~W.~M.} \bibnamefont{Bastiaansen}},
  \bibinfo{author}{\bibfnamefont{D.~K.~G.} \bibnamefont{de~Boer}},
  \bibnamefont{and} \bibinfo{author}{\bibfnamefont{D.~J.} \bibnamefont{Broer}},
  \bibinfo{journal}{Appl.\ Phys.\ Lett.} \textbf{\bibinfo{volume}{85}},
  \bibinfo{pages}{230} (\bibinfo{year}{2004}).

\bibitem[{\citenamefont{Syed et~al.}(2005)\citenamefont{Syed, Carbone,
  Rosenblatt, and Wen}}]{syed05}
\bibinfo{author}{\bibfnamefont{I.~M.} \bibnamefont{Syed}},
  \bibinfo{author}{\bibfnamefont{G.}~\bibnamefont{Carbone}},
  \bibinfo{author}{\bibfnamefont{C.}~\bibnamefont{Rosenblatt}},
  \bibnamefont{and} \bibinfo{author}{\bibfnamefont{B.}~\bibnamefont{Wen}},
  \bibinfo{journal}{J. Appl.\ Phys.} \textbf{\bibinfo{volume}{98}},
  \bibinfo{pages}{034303} (\bibinfo{year}{2005}).

\bibitem[{\citenamefont{Gibson et~al.}(2004)\citenamefont{Gibson, Courtial,
  Padgett, Vasnetsov, Pas'ko, Barnett, and Franke-Arnold}}]{gibson04}
\bibinfo{author}{\bibfnamefont{G.}~\bibnamefont{Gibson}},
  \bibinfo{author}{\bibfnamefont{J.}~\bibnamefont{Courtial}},
  \bibinfo{author}{\bibfnamefont{M.~J.} \bibnamefont{Padgett}},
  \bibinfo{author}{\bibfnamefont{M.}~\bibnamefont{Vasnetsov}},
  \bibinfo{author}{\bibfnamefont{V.}~\bibnamefont{Pas'ko}},
  \bibinfo{author}{\bibfnamefont{S.~M.} \bibnamefont{Barnett}},
  \bibnamefont{and}
  \bibinfo{author}{\bibfnamefont{S.}~\bibnamefont{Franke-Arnold}},
  \bibinfo{journal}{Opt.\ Express} \textbf{\bibinfo{volume}{12}},
  \bibinfo{pages}{5448} (\bibinfo{year}{2004}).

\bibitem[{\citenamefont{Leach et~al.}(2002)\citenamefont{Leach, Padgett,
  Barnett, Franke-Arnold, and Courtial}}]{leach02}
\bibinfo{author}{\bibfnamefont{J.}~\bibnamefont{Leach}},
  \bibinfo{author}{\bibfnamefont{M.~J.} \bibnamefont{Padgett}},
  \bibinfo{author}{\bibfnamefont{S.~M.} \bibnamefont{Barnett}},
  \bibinfo{author}{\bibfnamefont{S.}~\bibnamefont{Franke-Arnold}},
  \bibnamefont{and} \bibinfo{author}{\bibfnamefont{J.}~\bibnamefont{Courtial}},
  \bibinfo{journal}{Phys.\ Rev.\ Lett.} \textbf{\bibinfo{volume}{88}},
  \bibinfo{pages}{257901} (\bibinfo{year}{2002}).

\bibitem[{\citenamefont{Vaziri et~al.}(2002)\citenamefont{Vaziri, Weihs, and
  Zeilinger}}]{vaziri02}
\bibinfo{author}{\bibfnamefont{A.}~\bibnamefont{Vaziri}},
  \bibinfo{author}{\bibfnamefont{G.}~\bibnamefont{Weihs}}, \bibnamefont{and}
  \bibinfo{author}{\bibfnamefont{A.}~\bibnamefont{Zeilinger}},
  \bibinfo{journal}{Phys.\ Rev.\ Lett.} \textbf{\bibinfo{volume}{89}},
  \bibinfo{pages}{240401} (\bibinfo{year}{2002}).

\bibitem[{\citenamefont{Schadt et~al.}(1996)\citenamefont{Schadt, Seiberle, and
  Schuster}}]{schadt96}
\bibinfo{author}{\bibfnamefont{M.}~\bibnamefont{Schadt}},
  \bibinfo{author}{\bibfnamefont{H.}~\bibnamefont{Seiberle}}, \bibnamefont{and}
  \bibinfo{author}{\bibfnamefont{A.}~\bibnamefont{Schuster}},
  \bibinfo{journal}{Nature} \textbf{\bibinfo{volume}{381}},
  \bibinfo{pages}{212} (\bibinfo{year}{1996}).

\bibitem[{\citenamefont{Fan et~al.}(2003)\citenamefont{Fan, Ren, and
  Wu}}]{fan03}
\bibinfo{author}{\bibfnamefont{Y.-H.} \bibnamefont{Fan}},
  \bibinfo{author}{\bibfnamefont{H.}~\bibnamefont{Ren}}, \bibnamefont{and}
  \bibinfo{author}{\bibfnamefont{S.-T.} \bibnamefont{Wu}},
  \bibinfo{journal}{Opt.\ Express} \textbf{\bibinfo{volume}{11}},
  \bibinfo{pages}{3080} (\bibinfo{year}{2003}).

\bibitem[{\citenamefont{Chen et~al.}(1995)\citenamefont{Chen, Bos, Bryant,
  Johnson, Jamal, and Kelly}}]{chen95}
\bibinfo{author}{\bibfnamefont{J.}~\bibnamefont{Chen}},
  \bibinfo{author}{\bibfnamefont{P.~J.} \bibnamefont{Bos}},
  \bibinfo{author}{\bibfnamefont{D.~R.} \bibnamefont{Bryant}},
  \bibinfo{author}{\bibfnamefont{D.~L.} \bibnamefont{Johnson}},
  \bibinfo{author}{\bibfnamefont{S.~H.} \bibnamefont{Jamal}}, \bibnamefont{and}
  \bibinfo{author}{\bibfnamefont{J.~R.} \bibnamefont{Kelly}},
  \bibinfo{journal}{Appl.\ Phys.\ Lett.} \textbf{\bibinfo{volume}{67}},
  \bibinfo{pages}{1990} (\bibinfo{year}{1995}).

\bibitem[{\citenamefont{Todorov et~al.}(1984)\citenamefont{Todorov, Nikolova,
  and Tomova}}]{todorov84}
\bibinfo{author}{\bibfnamefont{T.}~\bibnamefont{Todorov}},
  \bibinfo{author}{\bibfnamefont{L.}~\bibnamefont{Nikolova}}, \bibnamefont{and}
  \bibinfo{author}{\bibfnamefont{N.}~\bibnamefont{Tomova}},
  \bibinfo{journal}{Appl.\ Opt.} \textbf{\bibinfo{volume}{23}},
  \bibinfo{pages}{4309} (\bibinfo{year}{1984}).

\bibitem[{\citenamefont{Eich et~al.}(1987)\citenamefont{Eich, Wendorff, Peck,
  and Ringsdorf}}]{eich87}
\bibinfo{author}{\bibfnamefont{M.}~\bibnamefont{Eich}},
  \bibinfo{author}{\bibfnamefont{J.~H.} \bibnamefont{Wendorff}},
  \bibinfo{author}{\bibfnamefont{B.}~\bibnamefont{Peck}}, \bibnamefont{and}
  \bibinfo{author}{\bibfnamefont{H.}~\bibnamefont{Ringsdorf}},
  \bibinfo{journal}{Makromol.\ Chem.\ Rapid.\ Commun.}
  \textbf{\bibinfo{volume}{8}}, \bibinfo{pages}{59} (\bibinfo{year}{1987}).

\end{thebibliography}

\newpage

\end{document}